\documentclass[aps,twocolumn,prd,showpacs,showkeys,preprintnumbers,floatfix,longbibliography,notitlepage,nofootinbib]{revtex4-2}
\usepackage[colorlinks=true,breaklinks=true]{hyperref}
\usepackage[utf8]{inputenc}
\usepackage{url}
\usepackage{gensymb}
\usepackage{mathtools}
\usepackage[dvipsnames]{xcolor}
\definecolor{darkred}{rgb}{0.5,0,0}
\definecolor{darkblue}{rgb}{0,0,0.5}
\definecolor{firebrick}{rgb}{0.75,0.125,0.125}
\definecolor{darkgreen}{rgb}{0,0.5,0}
\hypersetup{urlcolor=darkblue,
	    citecolor=darkgreen,
	    linkcolor=firebrick}

\usepackage{orcidlink}
\usepackage{lipsum}
\usepackage{tikz}
\usetikzlibrary{shapes,snakes}

\long\def\exclude#1{}

\newcommand{\ie}{{i.e.}}

\newcommand{\Refe}{Ref.}

\newcommand{\orcid}[1]{\href{https://orcid.org/#1}{\includegraphics[width=10pt]{orcid.pdf}}}
\bibpunct{[}{]}{,}{n}{}{,}

\begin{document}

\title{{\sc Nature Reviews Physics, Year In Review 2023}\\[0.3em]The Milky Way shines in high-energy neutrinos}

\author{Mauricio Bustamante
\orcidlink{0000-0001-6923-0865}} 
\email{mbustamante@nbi.ku.dk}
\affiliation{Niels Bohr International Academy, Niels Bohr Institute, University of Copenhagen, 2100 Copenhagen, Denmark}

\date{\today}

\begin{abstract}
 The most energetic astrophysical sources in the Milky Way, cosmic accelerators capable of producing high-energy cosmic rays, have resisted discovery for over a century.  Up to now, astrophysicists sought these sources mainly by scouring the Galaxy for the gamma rays they are expected to emit.  In 2023, the IceCube Neutrino Observatory discovered high-energy neutrinos from the Milky Way, inaugurating a telltale stream of evidence of cosmic-ray production and interaction in the Galaxy.
\end{abstract}

\maketitle

\tikzstyle{mybox} = [draw=firebrick, fill=white, very thick,
    rectangle, rounded corners, inner sep=10pt, inner ysep=20pt]
\tikzstyle{fancytitle}=[fill=firebrick, text=white, rounded corners]

\begin{tikzpicture}
 \node[mybox] (box){
  \hspace*{-0.55cm}
  \begin{minipage}{0.91\columnwidth}
   \begin{itemize}
    \item
     In 2023, the IceCube Neutrino Observatory discovered the first high-energy neutrinos coming from the Milky Way, most of them with energies in the range of 1--100~TeV.
    \item
     The neutrinos were discovered in the form of a diffuse flux from the Galactic Plane, whose spatial distribution matches that of previously detected gamma rays from the Milky Way.
    \item
     The flux of high-energy neutrinos from the Galactic Plane constitutes 6--13\% of the all-sky high-energy neutrino flux first discovered by IceCube in 2013.
    \item
     The discovery of high-energy Galactic neutrinos points to the long-predicted existence of PeVatrons in the Milky Way, \ie, astrophysical accelerators of cosmic rays up to PeV-scale energies, though none has been found yet.
   \end{itemize}
  \end{minipage}
 };
 \node[fancytitle, right=10pt] at (box.north west) {Key advances};
\end{tikzpicture}

\bigskip
\smallskip

For more than a hundred years, the origin of high-energy cosmic rays, the most energetic known particles, has eluded scientists.  However, the discovery of high-energy neutrinos from the Milky Way by the IceCube Neutrino Observatory~\cite{IceCube:2023ame} may have brought us closer to an answer.


\section*{Mysterious origins}

Cosmic rays are electrically charged atomic nuclei---from protons to iron---of extraterrestrial origin, with kinetic energies that outclass particle accelerators.  The most energetic cosmic rays, from outside the Milky Way, have energies exceeding $10^{20}$~eV, or tens of Joules, a billion times higher than nuclei accelerated by the Large Hadron Collider.  At lower energies, up to a few times $10^{15}$~eV, or 1~PeV, cosmic rays likely originate primarily inside the Milky Way.  

These Galactic comic rays hold insight into the star-formation activity in the Milky Way, and into the identity and properties of the most energetic processes in our Galaxy.  They are not too rare---on average, each square-meter of Earth is hit by a handful of PeV-scale cosmic rays every year---and so cosmic-ray detectors such as CASA-MIA, HEGRA, IceTop, KASCADE, and Tibet have amassed a large number of detected events.  Nonetheless, the origin of Galactic cosmic rays remains unknown.

Galactic cosmic rays roam the Milky Way, trapped by its magnetic field, of $\mu$Gauss intensity, for $10^4$--$10^7$~years.  During this time, their trajectories are scrambled by magnetic deflections: by the time they are detected on Earth, cosmic rays no longer point back to their sources.  Instead, what can be measured on Earth is a diffuse flux of cosmic rays, made up of the contributions emitted by all unresolved cosmic-ray sources, the memory of the location of individual sources long-ago lost.

\begin{figure*}[t!]
 \centering
 \includegraphics[]{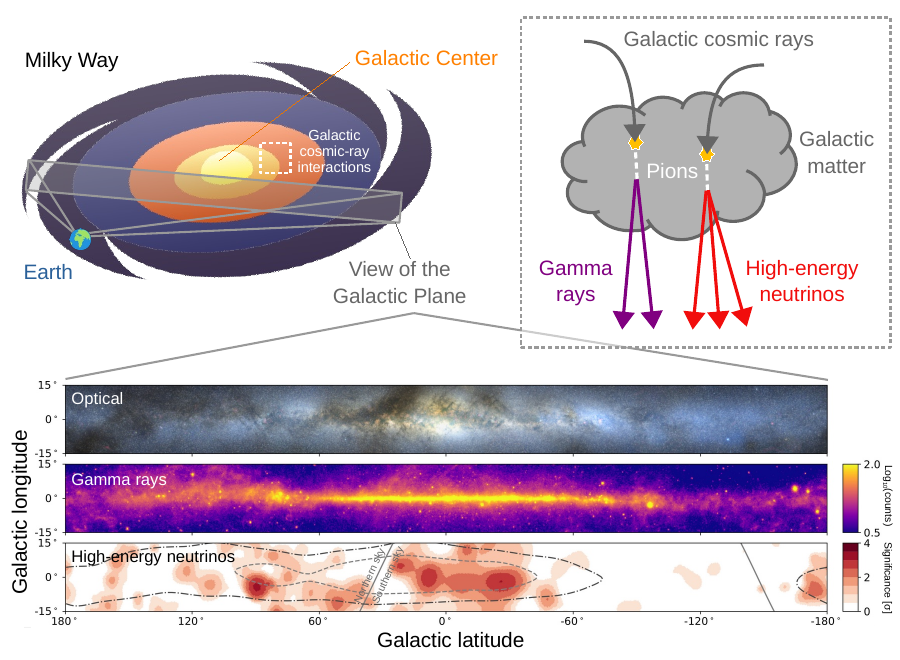}
 \caption{\textbf{\textit{The high-energy view of the Milky Way.}} High-energy Galactic cosmic rays, with PeV-scale energies (1~PeV $\equiv 10^{15}$~eV $\approx$ 0.2~mJ) are trapped by the magnetic field of the Milky Way, interact with surrounding matter, and generate high-energy gamma rays and neutrinos.  Viewed from Earth, most of the emission comes from the Galactic Plane, the edge-on view of the Galaxy, shown here in optical emission, gamma-ray photon counts detected by \textit{Fermi}-LAT, and high-energy neutrinos recently discovered by IceCube.  For neutrinos, the map is of significance of the detection over the background of atmospheric neutrinos and muons.  (Image credits: IceCube Collaboration, adapted from \Refe~\cite{IceCube:2023ame} / flaticon.com.)}
 \label{fig:diagram}
\end{figure*}

Although no source of PeV-scale cosmic rays---dubbed PeVatron---has been firmly identified within the Milky Way, or outside it, they must be sites that harbor the densities, speeds, and magnetic fields needed to confine and accelerate charged particles efficiently.  For Galactic cosmic rays, the most likely culprits are supernova remnants~\cite{Baade:1934}; particle acceleration might occur as their stellar ejecta sweeps up the interstellar medium.  


\section*{The gamma-ray picture}

During their long Galactic confinement, cosmic rays interact with matter and radiation in their path, near their sources or far from them.  These interactions create high-energy secondary particles: neutral and charged pions that decay into PeV gamma rays and neutrinos, respectively.  Unlike cosmic rays, gamma rays and neutrinos are not deflected by magnetic fields, and so their detected arrival directions can be used to aim true at the sites of cosmic-ray interactions in the Galaxy and to discover PeVatrons.  

So far, searches for Galactic PeVatrons have used primarily gamma rays.  The brightest structure in the sky in gamma rays is the Galactic Plane, the edge-on view of the Galaxy from our vantage point  26,000 light-years away from the Galactic Center.  Matter in the Galaxy, including potential PeVatrons, is concentrated along the Plane.  There, Galactic cosmic rays are confined, interact with their surroundings, and produce gamma rays.  The gamma-ray emission is brightest within a $10^\circ$ band around the Plane, and intensifies towards the Center, where most of the Galaxy's mass is packed.  This renders the Galactic Plane a candidate for high-energy neutrino emission, with a similar spatial distribution~\cite{Kheirandish:2020upj}.

By chance, the mean free path of PeV gamma rays to interact with cosmological photon backgrounds---chiefly, the cosmic microwave background---is about the same as our distance to the Galactic Center.  This means that many gamma rays created there with initial PeV energies will have lost most of their energy before reaching us with energies $10^3$--$10^6$ times lower, in the GeV--TeV range.  This clouds them as proxies of cosmic rays, since lower-energy gamma rays are also created by processes unrelated to cosmic rays, such as the synchrotron radiation from accelerated electrons.  Still, the detection of TeV gamma rays by modern observatories such as \textit{Fermi}-LAT, HAWC, H.E.S.S., MAGIC, and VERITAS have delivered hints of Milky Way PeVatrons; these were boosted by the long-awaited discovery of PeV gamma rays by LHAASO~\cite{LHAASO:2021gok}.


\section*{The neutrino picture}

High-energy neutrinos are not hampered by the same limitations as gamma rays.  Unlike gamma rays, neutrinos feel only the weak fundamental interaction.  As a result, they ordinarily disregard any matter and radiation in their path, and reach us with essentially the same energy as they were produced.  Thus, they provide unfettered information about their place and means of birth.  Yet, precisely because they interact weakly, they are notoriously harder to detect than gamma rays.  This is why Galactic high-energy neutrinos, long sought, were only recently discovered, even though the opening salvo of their story occurred a decade ago.

In 2013, the IceCube neutrino telescope discovered the first astrophysical neutrinos with PeV-scale energies, in the form of an all-sky diffuse flux~\cite{IceCube:2013low}.  Because the flux is largely isotropic, it is likely due predominantly to extragalactic neutrino sources distributed approximately homogeneously across the sky.  It took a few years for the first handful of candidate extragalactic sources to be identified~\cite{Kurahashi:2022utm}, and a full decade of observation for the sub-dominant contribution of neutrinos from the Galactic Plane to be resolved~\cite{IceCube:2023ame}.

IceCube is the largest neutrino telescope in operation: located at the South Pole, it instruments 1~km$^3$ of ice using photomultipliers buried over 1.5~km deep.  The extreme depth mitigates the background of cosmic rays, and of muons and neutrinos produced when they interact with the atmosphere, which may mimic the signals made by astrophysical neutrinos.  When a high-energy neutrino interacts with a proton or neutron in the ice, it triggers a particle shower whose light emission propagates through the ice and is detected by the photomultipliers.  From the spatial and temporal distributions of the light, the IceCube scientists infer the energy and arrival direction of the neutrinos.

Using these capabilities, the IceCube Collaboration reported~\cite{IceCube:2023ame} the discovery of high-energy neutrinos from the Galactic Plane with a robust statistical significance of nearly $4.5\sigma$.  Most of these neutrinos have energies between 1 and 100~TeV.  The search pivoted on previously observed gamma-ray emission from the Plane, by looking for neutrinos in the directions of known gamma-ray point sources and extended gamma-ray-emitting regions.  Although the search for point neutrino sources came seemingly just short of finding statistically significant evidence, the search for diffuse neutrino emission was successful.

The analysis used a spatial template of neutral-pion production along the Galactic Plane inferred from the spatial distribution of gamma rays from the Plane observed by \textit{Fermi}-LAT~\cite{Fermi-LAT:2012edv}, the distribution of gas in the Galaxy, and assumptions about the energy spectrum with which PeVatrons inject cosmic rays into the Galaxy.  The IceCube analysis assumed that charged-pion production follows the same template.  

The latter, upon decaying, yields a spatial template for high-energy neutrino emission that was contrasted against the directions of neutrinos detected by IceCube between May 2011 and May 2021.  The neutrino flux from the Galactic Plane is 6--13\% of the all-sky flux previously seen by IceCube, depending on the choice of template model, which roughly matches the most optimistic predictions~\cite{Ahlers:2015moa}. 

Spotting high-energy astrophysical neutrinos is an exercise in repeatedly finding the proverbial needle in the haystack, which in this case is the background of atmospheric muons and neutrinos created in the interactions of cosmic rays in the atmosphere.  IceCube detects the egregious rate of 100 million atmospheric muons for every astrophysical neutrino.  To combat this, searches for neutrino sources typically use `through-going tracks', events that reach IceCube primarily from the Northern Hemisphere, since atmospheric muons from that direction are absorbed during their passage through the Earth.  

However, most of the Galactic Plane, including the Galactic Center, is located in the Southern sky.  So, instead, IceCube resorted to using a different type of events, `cascades', from the Southern Hemisphere; they have worse angular resolution than tracks, but are affected by an order-of-magnitude smaller atmospheric neutrino background.  The key development that led to the discovery was the use of deep learning to identify more neutrino interactions in recorded data, with improved angular resolution.


\section*{Towards a full picture}

The discovery of neutrinos from the Galactic Plane is firm, but not without problems in its interpretation.  The aforementioned template of neutral-pion production predicts a neutrino flux that is five times smaller than the one observed.  A different template, the KRA$\gamma$ model~\cite{Gaggero:2015xza}, based on a more sophisticated treatment of cosmic-ray propagation in the Galaxy, instead predicts a flux 2--4 times higher than the one observed.  These mismatches may be rooted in a lack of freedom of these templates to accurately describe the energy spectrum of the detected neutrinos and might be addressed in follow-up analyses.

Ultimately, a full picture of the high-energy Milky Way will emerge from the combined study of cosmic rays, gamma rays, and neutrinos, and from understanding the interplay between them~\cite{Meszaros:2019xej, AlvesBatista:2021eeu}.  For neutrinos, IceCube has been joined by two neutrino telescopes in the Northern Hemisphere, currently under construction: KM3NeT in the Mediterranean Sea and Baikal-GVD in Lake Baikal, Russia, the successors of the long-running ANTARES and Baikal detectors.  Their joint observations, possibly with other planned neutrino telescopes~\cite{Ackermann:2022rqc} that include a potential IceCube upgrade, IceCube-Gen2, will sharpen our view of the high-energy Galactic sky in the coming decade.


\section*{Acknowledgments}

The author is grateful to Markus Ahlers and Steve Sclafani for their feedback on the manuscript, and is supported by the {\sc Villum Fonden} under project no.~29388.


\section*{Competing interests}

The author declares no competing interests.



%


\end{document}